\documentstyle[prl,aps,psfig]{revtex}

\begin{document}

\twocolumn[\hsize\textwidth\columnwidth\hsize\csname@twocolumnfalse\endcsname

\title{Thermodynamic stability of Fe/O \\
solid solution at inner-core conditions}
\author{D. Alf\`{e}$^\star$, 
G. D. Price$^\star$ and M. J. Gillan$^\dagger$ \medskip \\
$^\star$Research School of Geological and Geophysical
Sciences \\ Birkbeck College and University College London \\
Gower Street, London WC1E 6BT, UK \medskip \\
$^\dagger$Physics and Astronomy Department, University College London \\
Gower Street, London WC1E 6BT, UK}

\date{\today}

\maketitle

\begin{abstract}
We present a new technique which allows the fully {\em ab initio}
calculation of the chemical potential of a substitutional
impurity in a high-temperature crystal, including harmonic
and anharmonic lattice vibrations. The technique uses the combination
of thermodynamic integration and reference
models developed recently for the {\em ab initio} calculation of
the free energy of liquids and anharmonic solids. We apply the
technique to the case of the substitutional oxygen impurity in
h.c.p. iron under Earth's core conditions, which earlier
static {\em ab initio} calculations indicated to be
thermodynamically very unstable. Our results show that entropic
effects arising from the large vibrational amplitude of the oxygen
impurity give a major reduction of the oxygen chemical potential,
so that oxygen dissolved in h.c.p. iron
may be stabilised at concentrations up
a few mol~\% under core conditions.\\
~
\end{abstract}

]

The thermodynamic stability of oxygen dissolved in iron is a
key factor in considering the physics and chemistry of the Earth's
core. We present here a new technique which allows the {\em ab initio}
calculation of the chemical potential of an impurity in a
high-temperature solid solution, including harmonic
and anharmonic lattice vibrations. We report the application of
the technique to substitutional oxygen dissolved
in hexagonal close-packed (h.c.p.) iron
at Earth's core conditions, and we show that the Fe/O solid solution
is thermodynamically far more stable than expected from earlier
work. The new technique should find wide application to a range
of other earth-science problems.

It has long been recognised that the liquid outer core must
contain a substantial fraction of light impurities, since its
density is $6 - 10$~\% less than that estimated for pure
liquid Fe~\cite{birch64,poirier94};
similar arguments suggest that the inner core contains a smaller, but still
appreciable impurity fraction~\cite{stixrude97}.
The leading impurity candidates
are S, Si and O, and arguments have been presented for and
against each of them~\cite{poirier94}.
Ringwood~\cite{ringwood77} and
others~\cite{dubrovskiy72} have argued strongly
on grounds of geochemistry that oxygen must account for a large
part of the impurity content. However, it has proved difficult
to assess these ideas, because the Fe/O phase diagram is so poorly
known at Earth's core conditions. (For reference, we note that
the pressures at the core-mantle boundary, the inner-core boundary (ICB)
and the centre of the Earth are 136, 330 and 364~GPa respectively;
the temperatures at the core-mantle boundary and the ICB are poorly
established, but are believed to be in the region of 
4000 and 6000~K respectively.)

The thermodynamic stability of dissolved oxygen is governed by
the free energy change in the reaction
\begin{equation}
( n - 1 ) {\rm Fe (solid)} +
{\rm FeO (solid)} \rightarrow {\rm Fe}_n{\rm O (solid \; solution)}
\end{equation}
Let $\Delta G$ be the increase of Gibbs free energy as this reaction
goes from left to right, excluding the configurational contribution
associated with the randomness of the lattice sites occupied by dissolved
O. Then the maximum concentration (number of O atoms per crystal lattice
site) at which dissolved O is thermodynamically stable with respect to
precipitation of FeO is $c_{\rm max} = \exp ( - \Delta G / k_{\rm B} T )$.
Several years ago, Sherman~\cite{sherman95}
used {\em ab initio} calculations
based on density functional theory (DFT) to calculate the zero-temperature
limit of $\Delta G$, i.e. the enthalpy $\Delta H$ of reaction~(1).
He found that $\Delta H$ is very large ($\sim 5$~eV at the ICB pressure
of 330~GPa), 
and concluded that the concentration
of dissolved O in the inner core must be completely negligible.
His argument has been widely cited. However, these were
static, zero-temperature calculations, which entirely ignored entropic
effects. We shall show here that the high-temperature entropy of
dissolved O produces such a large reduction of free energy
that Sherman's argument should be treated
with caution when considering core temperatures.

Our {\em ab initio} calculations are based on the well established
DFT methods used in virtually all
{\em ab initio} investigations of solid
and liquid Fe~\cite{stixrude94,soderlind96,dewijs98a,alfe99a,alfe00a},
including Sherman's~\cite{sherman95}. We
employ the generalised gradient approximation
for exchange-correlation energy, as formulated by
Perdew {\em et al.}~\cite{perdew92},
which is known to give very accurate results for the low-pressure
elastic, vibrational and magnetic properties 
of body-centred cubic (b.c.c.) Fe, the
b.c.c.~$\rightarrow$ h.c.p. transition pressure, and the pressure-volume
relation for h.c.p. Fe up to over
300~GPa~\cite{stixrude94,alfe00a}. We use the
ultra-soft pseudopotential implementation~\cite{vanderbilt90}
of DFT with plane-wave
basis sets, an approach which has been demonstrated to give results
for solid Fe that are virtually identical to those of all-electron
DFT methods~\cite{alfe00a}.
Our calculations are performed using the VASP
code~\cite{kresse96},
which is exceptionally stable and efficient for metals. The
technical details of pseudopotentials, plane-wave cut-offs, etc.
are the same as in our previous work on the Fe/O system~\cite{alfe99b}.

\begin{figure}
\centerline{\psfig{figure=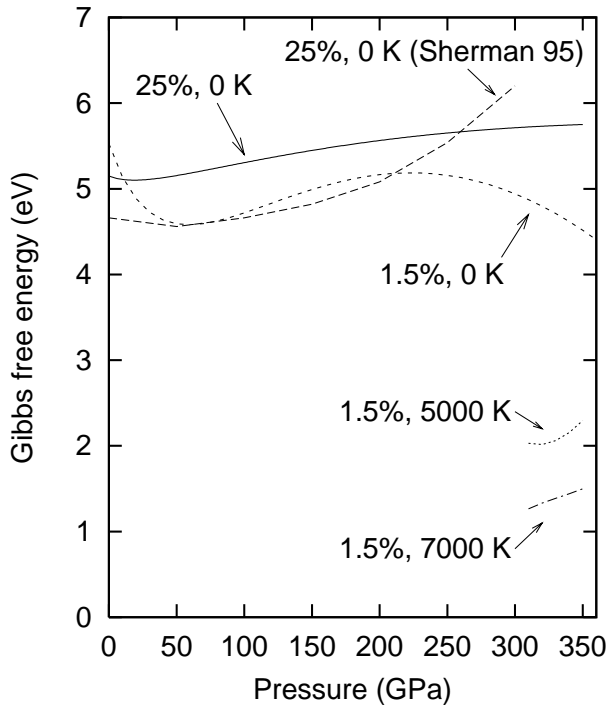,height=3.9in}}
\caption{Gibbs free energy $\Delta G$ of the reaction
$(n-1) {\rm Fe (solid)} + {\rm FeO (solid)} \rightarrow
{\rm Fe}_n{\rm O} \; \; {\rm (solid)}$
from {\em ab initio} calculations. Upper curves show
zero-temperature values (i.e. enthalpies) for 25~mol~\% concentration
from Refs.~\protect\cite{alfe99b} (solid curve) and \protect\cite{sherman95} (long
dashes) and for 1.5~\% concentration (short dashes) from present
work. Short lower curves show high-temperature results for
$\Delta G$ from present work at 5000~K (dots) and 7000~K (chain line).}
\end{figure}

We first report static zero-temperature results for the
enthalpy $\Delta H$ of the reaction~(1). Sherman's results~\cite{sherman95},
later confirmed by the present authors~\cite{alfe99b},
were obtained for the high
O concentration of 25~mol~\%, corresponding to $n = 3$, 
but here we wish to focus
on the dilute limit, and we take $n = 63$, which gives a mole fraction of
1.5~\%. To do this, we treat a 64-atom supercell with the h.c.p.
structure containing a single O substitutional, and we calculate
the total ground-state energy and pressure for a sequence of
atomic volumes, with all atoms relaxed to their equilibrium
positions at each volume. The enthalpy of the pure iron system
is obtained from total-energy and pressure calculations for
a single unit cell of the h.c.p. crystal. For the FeO crystal,
we obtain the enthalpy from total-energy and pressure calculations
on a unit cell of the NiAs structure. (The high-pressure
stable structure of FeO is believed to be either NiAs or inverse-NiAs;
the relative stability of the two structures 
has been controversial~\cite{fei94,cohen97,fang99}, 
but our own {\em ab initio}
calculations indicate that the NiAs structure is slightly
more stable at pressures above {\em ca.}~145~GPa.)
The enthalpy $\Delta H$ is
reported as a function of pressure in Fig.~1, where we
also show Sherman's results and our own for the 25~mol~\%
case. We see that $\Delta H$ at 1.5 mol~\% is between 0.5 and 1.0~eV
lower than at 25 mol~\%, having a value of {\em ca.}~4.7~eV at
330~GPa, but this is still very large and Sherman's arguments would
remain valid if this represented a good estimate of the free energy of
reaction~(1) at Earth's core temperatures.

\begin{figure}
\centerline{\psfig{figure=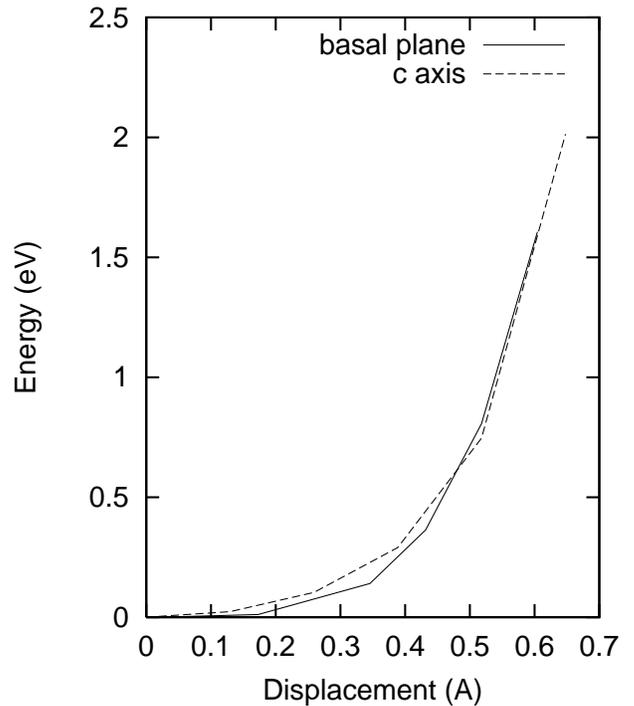,height=3.9in}}
\caption{Calculated energy (eV units) as function of
displacement (\AA\ units) of O impurity from its equilibrium
site in basal plane (solid curve) and along $c$-axis (dashed curve).
Results are for crystal volume of 6.97~\AA$^3$/atom.}
\end{figure}

To get an idea of the freedom of movement of the substitutional
O atom, and hence its vibrational entropy, we now perform a series
of calculations in which the O atom is displaced by different
amounts from its equilibrium site, with all other atoms held fixed
at the equilibrium positions they have when the O atom is at
its own equilibrium position. Results for the energy variation
with displacement along the $c$-axis and in the basal plane for the
crystal volume of 6.97~\AA$^3$/atom are shown in
Fig.~2. Since we are interested in core temperatures, we also
mark the energy $k_{\rm B} T$ for $T = 6000$~K. We see that
in the energy range set by this $T$ the energy surface is extremely
anharmonic, with almost vanishing curvature at the equilibrium site
and large curvatures for large displacements. Clearly, the vibrational
root-mean-square (r.m.s.) O displacement will be $\sim 0.45$~\AA\
(our direct {\em ab initio} molecular-dynamics calculations
confirm this value).
For comparison, we estimate the r.m.s. O displacement in FeO at
the same $P$ and $T$ to be $\sim 0.23$~\AA. This implies
a substantial vibrational entropy for the O impurity,
because it fits so loosely into the Fe lattice. For the same reason, the 
vibrations of the 12 Fe neighbours of the O impurity will be softened,
and this will also increase the entropy.

To make quantitative statements about high-temperature behaviour,
we need to calculate the {\em ab initio} Gibbs free energies, rather than
zero-temperature enthalpies. 
The {\em ab initio} calculation of the Helmholtz free energies $F$
of the Fe and FeO perfect crystals is straightforward
in the harmonic approximation, since this requires only
the energy of the static lattice and the {\em ab initio}
lattice vibrational frequencies, which we calculate by the
small-displacement method, as discussed in several previous
papers~\cite{alfe99a,kresse95,vocadlo99,vocadlo00}.
From $F$, we then directly obtain the Gibbs free energy $G = F + PV$,
by calculating the pressure $P$ as $- ( \partial F / \partial V )_T$,
with $V$ the volume. The only difficult part
of the present problem is therefore the calculation of $F$ (and hence $G$)
for the ${\rm Fe}_n {\rm O}$ crystal containing the substitutional
O atom.
This free energy must include
the vibrations of many shells of neighbours of the O impurity. The
harmonic approximation will clearly not suffice. We meet this
challenge by drawing on recently developed {\em ab initio} methods for 
calculating the free energies of liquids and anharmonic
solids~\cite{alfe99a,dewijs98b,alfe00b}.
These methods rely on two things: empirical reference models, parameterised
to accurately mimic the {\em ab initio} energies; and the
technique of `thermodynamic integration', used to determine
free energy differences. Our overall strategy will be to obtain the {\em ab
initio} free energy $F_{\rm Fe/O}^{\rm AI}$ of the O-substitutional
system by starting from the {\em ab initio} free energy
$F_{\rm Fe}^{\rm AI}$ of pure Fe and using thermodynamic integration
to compute the free energy change $F_{\rm Fe/O}^{\rm AI} -
F_{\rm Fe}^{\rm AI}$ that results from converting a single Fe
atom into an O atom. 

We recall briefly that thermodynamic integration~\cite{frenkel96}
is a general
technique for calculating the difference of free energies
$F_1 - F_0$ of two systems containing the same number $N$ of
atoms but having different total-energy functions
$U_0 ( {\bf r}_1 , {\bf r}_2 , \ldots {\bf r}_N )$ and
$U_1 ( {\bf r}_1 , {\bf r}_2 , \ldots {\bf r}_N )$, with
${\bf r}_i (i = 1 , 2 \ldots N)$ the atomic positions. The technique
relies on the equivalence between the free energy difference
and the reversible work done on switching the total energy 
function continuously from $U_0$ to $U_1$. The work done is:
\begin{equation}
F_1 - F_0 = \int_0^1 d \lambda \, \langle U_1 - U_0 \rangle_\lambda \; ,
\end{equation}
where the thermal average $\langle \, \cdot \, \rangle_\lambda$
is evaluated in the canonical ensemble generated by the switched
energy function $U_\lambda$
defined by:
\begin{equation}
U_\lambda = ( 1 - \lambda ) U_0 + \lambda U_1 \; .
\end{equation}
To apply this in practice, we use molecular dynamics simulation
to evaluate the average $\langle U_1 - U_0 \rangle_\lambda$
at a sequence of $\lambda$ values and we perform the integration
over $\lambda$ numerically.

In principle, we could calculate $F_{\rm Fe/O}^{\rm AI} -
F_{\rm Fe}^{\rm AI}$ by identifying $U_0$ and $U_1$ as the {\em ab initio}
total energy functions $U_{\rm Fe}^{\rm AI}$ and
$U_{\rm Fe/O}^{\rm AI}$ of the pure-Fe and O-substitutional
systems, but this brute-force approach is computationally
prohibitive at present. It is also unnecessary, since exactly the same result
can be achieved much more cheaply by using empirical reference
models.
In our recent work on liquid Fe~\cite{alfe00a}, we have
found that a simple inverse-power pair potential $\phi ( r ) =
A / r^\alpha$ reproduces the {\em ab initio} total energy very
accurately; for the anharmonic high-temperature Fe crystal,
a linear combination of this pair-potential model with an {\em ab initio}
harmonic description has been very effective~\cite{alfe99a}. We denote the
total energy of this latter anharmonic model by
$U_{\rm Fe}^{\rm ref} ( {\bf r}_1 , \ldots {\bf r}_N )$, where
${\bf r}_i$ are the atomic positions.

In order to make a reference system for the Fe/O system containing
a single substitutional O atom, whose total energy is 
$U_{\rm Fe/O}^{\rm ref} ( {\bf r}_1 , \ldots {\bf r}_N )$ (${\bf r}_1$
is the position of the O atom), we simply delete all terms
in $U_{\rm Fe}^{\rm ref}$ involving atom 1 and replace them
with a pair interaction potential $\chi ( r ) =
B / r^\beta$. All parts of $U_{\rm Fe/O}^{\rm ref}$ involving
the $N - 1$ Fe atoms $2, 3, \ldots N$ remain precisely as they
are in $U_{\rm Fe}^{\rm ref}$. 
Our procedure for determining $B$ and $\beta$ starts by requiring
that the pair potential $\chi (r)$ should reproduce as well as
possible the dependence of the total energy on position of
the O atom with all Fe atoms held fixed, i.e. the curves shown
in Fig.~2. The values for $B$ and $\beta$ thus obtained give us
an initial form for $U_{\rm Fe/O}^{\rm ref}$. This initial
$U_{\rm Fe/O}^{\rm ref}$ is then used in a classical
MD simulation to generate a long trajectory at the temperature
of interest, from which we take 100 statistically independent
configurations. The full {\em ab initio} energies are calculated
for these configurations, and the $B$ and $\beta$ parameters
are readjusted to give a least squares fit to these energies.
Finally, the $U_{\rm Fe/O}^{\rm ref}$ obtained from the new
$B$, $\beta$ is used to generate a further 100 statistically
independent configurations, and $B$ and $\beta$ are adjusted once
more to fit the {\em ab initio} energies of these configurations.
The $B$ and $\beta$ produced by this final step are found to be
essentially identical to those in the previous step, and we accept them as
the optimal values for the assumed form of $\chi (r)$.

The free energy difference $F_{\rm Fe/O}^{\rm AI} -
F_{\rm Fe}^{\rm AI}$ between the Fe/O and pure Fe systems is now
expressed as the sum of three contributions: the difference
$F_{\rm Fe/O}^{\rm ref} - F_{\rm Fe}^{\rm ref}$ between
the reference systems, and the two differences
$F_{\rm Fe/O}^{\rm AI} - F_{\rm Fe/O}^{\rm ref}$ and
$F_{\rm Fe}^{\rm ref} - F_{\rm Fe}^{\rm AI}$ between the {\em ab initio}
and reference systems. All three of these differences are calculated
by thermodynamic integration. We emphasise that, although
the reference systems play a vital role, the final result
for $F_{\rm Fe/O}^{\rm AI} - F_{\rm Fe}^{\rm AI}$ does
not depend on how they are chosen.

Thermodynamic integration to get $F_{\rm Fe/O}^{\rm ref} -
F_{\rm Fe}^{\rm ref}$ is easy and rapid, since only simple
potential models are involved. We have tested size effects by using
simulated systems containing up to 768 atoms, but we find that
with only 288 atoms the size-effect errors are less than
the statistical errors of {\em ca.}~30~meV. For the
{\em ab-initio}/reference differences
$F_{\rm Fe/O}^{\rm AI} - F_{\rm Fe/O}^{\rm ref}$ and
$F_{\rm Fe}^{\rm AI} - F_{\rm Fe}^{\rm ref}$, the fluctuations
of $U_{\rm Fe/O}^{\rm AI} - U_{\rm Fe/O}^{\rm ref}$
and $U_{\rm Fe}^{\rm AI} - U_{\rm Fe}^{\rm ref}$
are so small that explicit thermodynamic integration over
$\lambda$ is unnecessary, and we can use instead the small-$\lambda$
approximation explained elsewhere~\cite{small}. We have studied the size
errors for these {\em ab initio}/reference differences, and
we find that results obtained with 36, 64 and 96 atoms are identical
within statistical errors. The overall statistical error
on the {\em ab initio} difference $F_{\rm Fe/O}^{\rm AI} -
F_{\rm Fe}^{\rm AI}$ is {\em ca.}~90~meV. We have repeated
all the above calculations at the four
volumes 6.86, 6.97, 7.20 and 7.40~\AA$^3$/atom, and from
the dependence on volume we obtain the pressure change
on replacing Fe by O and hence the Gibbs free energy difference
$G_{\rm Fe/O}^{\rm AI} - G_{\rm Fe}^{\rm AI}$.

Our calculated Gibbs free energies $\Delta G$ for reaction~(1) 
are displayed in Fig.~1 for the two temperatures 5000 and 7000~K. We
note the very large entropic lowering of $\Delta G$, which, at
$P = 330$~GPa comes down from 4.7 to {\em ca.}~1.7~eV at the temperature
$T \simeq 6000$~K expected at the ICB. This is still
a substantial positive value, but implies that the
stability-limit concentration $c_{\rm max} = \exp ( - \Delta G /
k_{\rm B} T )$ is {\em ca.}~3~mol~\%, which is far from negligible.

In assessing our $c_{\rm max}$ value, one should note the remaining
uncertainties in our calculations. First, we have ignored anharmonicity
in the pure Fe and FeO crystals. Our recent work on the effect
of anharmonicity in pure Fe~\cite{small} 
showed that at the melting point, anharmonicity
can contribute as much as 70~meV/atom to the free energy;
the same might be true of FeO. These effects could shift
$\Delta G$ by perhaps 0.15~eV. Second, there is the question of strong
electronic correlation in FeO, which is a prototypical Mott insulator
at low pressures. Such correlation effects will be much weakened
at Earth's core pressures, but could still shift $\Delta G$ by
a few tenths of an~eV. This means that our prediction for $c_{\rm max}$
at a given temperature is probably not reliable to better than a
factor of 3. We are therefore cautious about the detailed numerical
value of $c_{\rm max}$, and claim only that it could be a few
mol~\% at ICB pressure and temperature. 

In summary, we conclude that, because
substitutional oxygen atom
fits so loosely into the Fe lattice and has so much
freedom of movement, it undergoes
a very large entropic lowering
of free energy at high temperatures, this lowering being as much
as 3~eV at 6000~K and 330~GPa. The consequence is that
substitutional O dissolved in h.c.p. Fe may be thermodynamically 
stabilised at concentrations
up to a few mol~\%. Earlier {\em ab initio}
calculations~\cite{sherman95}
which ignored entropic effects should therefore not be taken
at face value.
Finally, we point out that
a wide range of geological problems depend on an understanding
of chemical potentials -- for example, all problems concerned
with the partitioning of elements between coexisting phases.
The {\em ab initio} techniques for calculating chemical
potentials outlined here should therefore be of wide interest.
\medskip

\noindent
{\bf Acknowledgments.} The work of D.A. was supported by
NERC Grant No. GR3/12083. Allocations of time on the Cray T3E
machines at the Manchester CSAR service and at Edinburgh Parallel
Computer Centre were provided by the UK Car-Parrinello Consortium
(EPSRC grant GR/M01753) and the Minerals Physics Consortium (NERC grant
GST/02/1002). Some of the calculations were performed at the UCL
HiPerSPACE Centre, partially funded by the Joint Research
Equipment Initiative.

\end{document}